\documentclass{JAC2003}

\usepackage{array}
\usepackage{graphicx}
\usepackage{booktabs}

\begin{document}

\title{DARK CURRENT STUDIES FOR SWISSFEL}

\author{F.~Le Pimpec, R.~Zennaro, S.~Reiche, E.~Hohmann, A.~Citterio, A.~Adelmann
\\ Paul Scherrer Institut, 5232 Villigen, Switzerland
\\B.~Grigoryan
\\ CANDLE Yerevan, Armenia}

\maketitle

\begin{abstract}

Activation of the surroundings of an accelerator must be quantified
and those data provided to the official agencies. This is a
necessary step in obtaining the authorization to operate such an
accelerator. SwissFEL, being a 4$^{th}$ generation light source,
will produce more accelerated charges, which are dumped or lost,
than conventional 3$^{rd}$ generation light source, such as the
Swiss Light Source. We have simulated the propagation of a dark
current beam produced in the photoelectron gun using tracking codes
like ASTRA and Elegant for the current layout of SwissFEL.
Experimental studies have been carried out at the SwissFEL test
facilities at PSI (C-Band RF Stand and SwissFEL Injector Test
Facility), in order to provide necessary input data for detailed
study of components (RF gun and C-band RF structures) using the
simulation code OPAL. A summary of these studies are presented.

\end{abstract}

\section{DARK CURRENT SIMULATIONS FOR SWISSFEL}

In an RF gun, the dark current is initially generated by field
emission from the photocathode and around the irises of the cavities
for all accelerating RF structures. The direction of propagation of
the electrons obviously depends on the RF phase and the efficiency
of propagation depends of the type of cavity, traveling or standing
wave (TW, SW). Impinging electrons to the surrounding walls of the
cavity can produce secondaries. Those secondary electrons can also
be captured and transported along the beam line and add to the dark
current.

\subsection{ASTRA simulations}

\begin{figure}[htb]
   \centering
   \includegraphics*[width=\columnwidth]{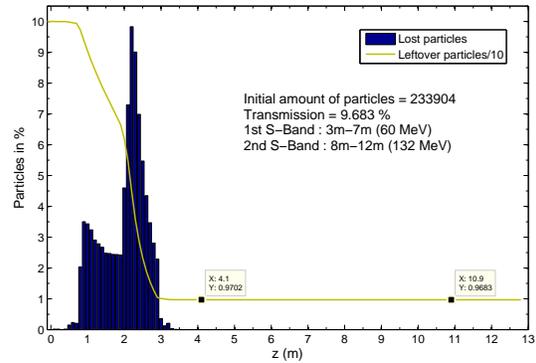}
   \caption{Histogram of the particle losses at the walls of the accelerator up to the end
   of the second S-band cavity (132~MeV). The continuous line represents the surviving particles
   in percent (/10) of the initial remaining particles.}
   \label{figlossesASTRA}
\end{figure}

We have simulated the dark current of the SwissFEL
\cite{SwissFEL:CDR} S-band standing wave photogun to the end of the
second 4~m long S-band traveling wave structure, by using the
tracking code ASTRA \cite{ASTRA}. The initial electron bunch was
produced by using the SwissFEL nominal bunch, spreading its
dimension in time and space. The emission time was limited to 120~ps
which only covers one RF bucket $\pm$60$^\circ$ around the on-crest
phase. We turned off the space charge option. The dark current is in
reality emitted at some threshold and at every RF bucket. This
simple approximation is sufficient as every other dark current
bucket propagating downstream of the RF gun will be transported
identically by the machine optics. From the initial 300~k particles
of the bunch, $\sim$22\% are lost on the cathode. The losses on the
walls (in percent) are quantified along the first 13~m of the
machine, Fig.\ref{figlossesASTRA}, using the remaining $\sim$234~k
particles. Only 9.7\% of those remaining particles reach the
beginning of the third S-band accelerating structures. As shown in
Fig.\ref{figlossesASTRA}, more than 90\% of the bunch is lost before
even reaching the first S-band structure with a $\sim$7~MeV kinetic
energy. The 9.7\% remaining particles are concentrated in the core
of the initial bunch. The output distribution was reused as an input
for the elegant tracker \cite{elegant} starting from the beginning
of the third S-band structure.

\subsection{Elegant simulations}

\begin{figure}[htb]
   \centering
   \includegraphics*[width=\columnwidth,clip=]{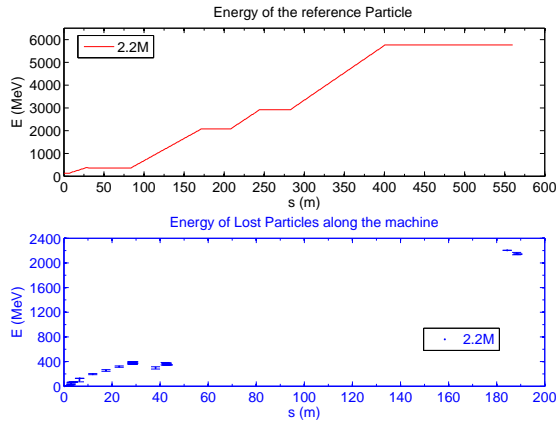}
   \caption{Top: Energy gained by the reference particle up to the end of the Aramis beam line.
    Bottom: Average energy and the standard deviation of the energy of the lost particles along the machine.}
   \label{figElosses}
\end{figure}

\begin{figure}[htb]
   \centering
   \includegraphics*[width=\columnwidth,clip=]{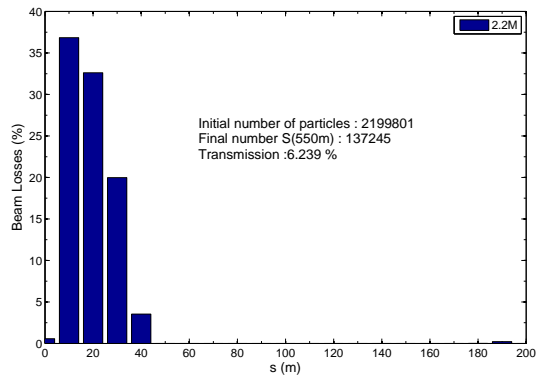}
   \caption{Histogram of the particle losses at the walls of the accelerator up to the end
   of the Aramis beam line (6~GeV)}
   \label{figlossesElegant}
\end{figure}

The leftover particles from the ASTRA simulation were not numerous
enough to do a proper elegant simulation. We have multiplied the
input distribution by cloning every particle with a small change in
their positions and momenta. We hence produced 2.2 million
particles. The tracking was done using the SwissFEL Elegant model
which includes the physical apertures of the machine and by adding
collimators toward the end of the Aramis beam line
\cite{SwissFEL:CDR}. The space charge and wake field options were
turned off. Fig.\ref{figElosses} (bottom plot) and
Fig.\ref{figlossesElegant} show the location where  particles are
lost and with which energy and energy spread. The energy spread for
the lost particles is small, less than 1\%, as can be seen by the
small error bars displayed in the bottom plot of
Fig.\ref{figElosses}. Losses are concentrated before the first bunch
compressor (s$\sim$50~m) and at the second bunch compressor
(s~$\sim$~200~m), as shown in Fig.\ref{figlossesElegant}. No other
losses are recorded after the second bunch compressor. The
transverse collimators placed at the end of the machine
(s~$>$~400~m) induce dark current losses only for apertures smaller
than 2~mm in radius.

Further simulations were carried out to find out the effects of
different acceleration schemes on losses, due to potential SwissFEL
operation conditions. The case presented in Fig.\ref{figElosses} and
\ref{figlossesElegant} uses the full acceleration in LINAC~3 up to
$\sim$6~GeV. LINAC~3 is the last stage of acceleration for the hard
X-ray beamline. It raises the beam energy from $\sim$3~GeV to 6~GeV.
This linac starts downstream of the beamline switchyard at s
$\approx$ 280~m. More details about the layout of the machine can be
found in reference \cite{SwissFEL:CDR}. The other cases schemes
simulated were the following.
\begin{itemize}
 \item[*] Acceleration in LINAC~2 to 3~GeV and deceleration to
 2.1~GeV before the entrance in LINAC~3.
 \item[*] Acceleration in LINAC~2 to 3~GeV and deceleration to
 2.1~GeV using the last C-band structures at the end of LINAC 3.
 \item[*] Acceleration in LINAC~3 to 3.5~GeV and deceleration to
 2.1~GeV using the last C-band structures at the end of LINAC 3.
\end{itemize}
They show no differences in losses for the same aperture of the
collimators as the case presented here. Small differences arose for
collimator sizes below the 2~mm radius.

Finally, we used elegant to track back some potential dark current
produced by the TW C-band RF structures \cite{Zennaro:IPAC2012}. In
a TW structure the RF capture phase exists only in the direction of
propagation of the RF wave. Due to the design of our C-band
structure, we do not expect any dark current above a few MeV
traveling upstream of the LINAC. In elegant reversing the linac
model will also reverse the direction of propagation of the RF hence
allowing capture of an upstream propagating electron beam. Despite
this limitation, we find that no dark current produced by the last
C-band structure of LINAC~3 makes it back to the beginning of the
machine for an initial electron energy below 25~MeV. This energy
exceed by far the energy gained by electrons accelerated through 2
or 3 cells in a structure.

\subsection{OPAL simulations}

\begin{figure}[htb]
   \centering
   \includegraphics*[width=\columnwidth, clip=]{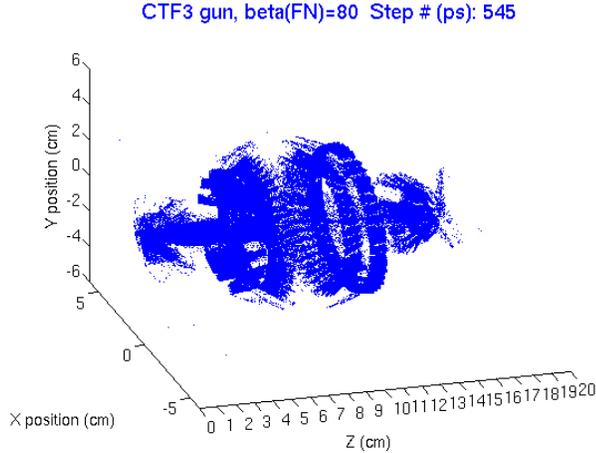}
   \caption{Propagation of the dark current in a SW photogun.
   The electron propagates upstream toward the cathode and downstream toward the exit of the gun}
   \label{figCTFgun545}
\end{figure}

The preceding simulations have been carried out by using a blown up
bunch based on the initial SwissFEL bunch. In order to use a more
adequate bunch distribution, we have used the dark current
simulation capabilities of OPAL \cite{Adelm:2008,Adelm:2010},
turning off the secondary electron emission switch. We have tested
the emission and propagation of a dark current beam using a
Fowler-Nordheim (FN) threshold for emission of 40~MV/m and a FN
enhancement factor $\beta$ of 80. For the component simulation, we
have used the actual layout of the SwissFEL injector test facility
gun, labeled CTF3 in Fig.\ref{figCTFgun545} \cite{250Injector:CDR}.
The CTF3 gun is a two and a half cell SW RF structure. The RF peak
field at t=0~ps is 100~MV/m on axis. For every picosecond the RF
phase changes by 1$^\circ$. The electrons are created at first on
the cathode side and on the downstream side of the irises and
electrons start propagating downstream of the beam line. 180$^\circ$
later the electrons are created in the upstream side of the irises
and start propagating upstream (toward the cathode).
Fig.\ref{figCTFgun545} shows a snapshot at t=545~ps of the
propagation of the electrons. The 6D phase space output of the
simulation can be used as an input for the OPAL or elegant trackers.
The modified module also handles TW structures. We plan to simulate
the dark current production and propagation from an actual C-band
test structure \cite{Zennaro:IPAC2012}.

\section{EXPERIMENTAL RESULTS AT VARIOUS SWISSFEL FACILITIES}

\subsection{Dark current at the SwissFEL injector}

\begin{figure}[htb]
   \centering
   \includegraphics*[width=\columnwidth, clip=]{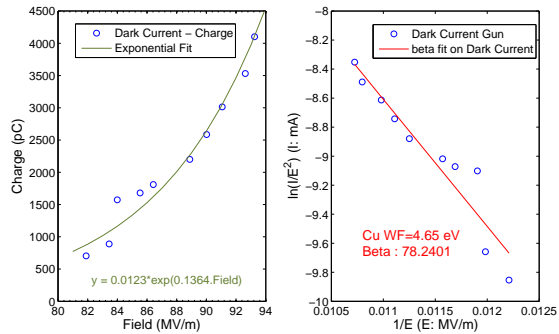}
   \caption{Left: CTF3 gun charge emission vs accelerating field. Right:
   Enhancement factor $\beta$ determination.}
   \label{figCTF3FN}
\end{figure}

In order to provide proper inputs for the simulation we measured the
dark current produced by the injector RF photogun (CTF3), and by the
4~m long S-band TW structure. Its propagation along the beamline was
also studied \cite{250Injector:CDR}. The CTF3 gun produces around
6~nC of dark current at nominal power (100~MV/m accelerating
gradient). The charge is measured using the wall current monitor
(WCM) located downstream of the gun. A second WCM is available
downstream of the bunch compressor (BC). The emission of dark
current in the gun is FN driven, as shown by the left plot in
Fig.\ref{figCTF3FN}. Using equation~\ref{EquFNbeta} in
\cite{lepimpec:2011}, we have extracted the FN enhancement factor
$\beta$ = 78 for such structure, Fig.\ref{figCTF3FN} (right plot),

\begin{equation}
\beta = \frac{6.83 \cdot 10^3 \times \Phi^{\frac{3}{2}}}{k_2}
\label{EquFNbeta}
\end{equation}

where $\Phi =4.65~eV$ is the work function, and k$_2$ is the slope
of the fit. Warm RF cavities with $\beta$ of $\sim$50 are usually
consider "good" structures.

In order to determine the energy spread of the dark current we have
used two techniques. In one case, we have varied the input power at
the gun and looked at the beam at the screen located at the dump of
the energy spectrometer. In the other case, we swept the beam on the
screen using the spectrometer dipole at nominal power. We measured a
threshold for dark current at 4.3~MeV and a maximum energy of
7.2~MeV. The dark current energy on the lowest end could be lower
than the measured 4.3~MeV, but we could not detect any electrons
neither on the WCM nor on any scintillating screens of the beamline.
The dark current emitted by the gun was measured, by the use of a
WCM, to be 6~nC increasing by a couple of nC over time (months). The
transmission of this dark current through the injector accelerator
(5\%) is in agreement with elegant simulations (6\%,
Fig.\ref{figlossesElegant}).

We have tried to observe the dark current emitted solely by the TW
S-band structures, running at 28~MW of input power (17~MV/m), on the
upstream and downstream screens of each structures. No dark current
was detected. A gamma ray spectrometer, facing the second S-band
structure at a distance of 1~m and then 30~cm, showed no signal
during the S-band operation. From these series of experiments, we
concluded that the dark current measured downstream of the bunch
compressor (300~pC) came solely from the gun dark current.

\begin{figure}[htb]
   \centering
   \includegraphics*[width=\columnwidth, clip=]{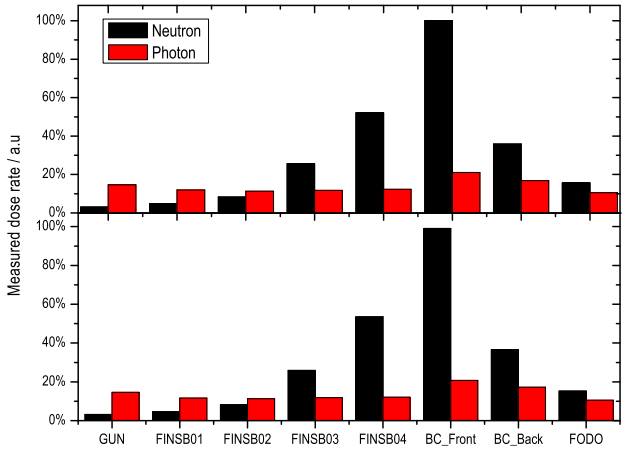}
   \caption{Measured radiation dose rate in arbitrary units along the machine measured 1~m away from each component.
   The dose rate is normalized to the highest measured value. Above : Laser is ON and all RF components are ON.
   Below : Laser is OFF and all RF components are ON}
   \label{figRadMeasur}
\end{figure}

Off-line radiation measurements performed at the injector were in
qualitative agreement with the ASTRA results. \newline On-line
radiation measurements show a significant high dose rate (normally
given in terms of ambient dose equivalent) at the entrance of the
fourth TW S-band structure (FINSB04), instead of the first one, and
at the entrance of the BC (BC\_Front), Fig.\ref{figRadMeasur}. This
seems contradictory to the simulations results, but one has to take
into consideration the energy of the particles at FINSB04 and at the
entrance of the first S-band structure (FINSB01). More electrons are
lost at 7~MeV, entrance of FINSB01, than at 200~MeV, entrance of
FINSB04, but the radiation production for electrons with an energy
of 200~MeV is higher than at 7~MeV. Photon and neutron dose rate (in
arbitrary units) is produced by the losses of the dark current on
the accelerator chamber wall, Fig.\ref{figRadMeasur}. The results
are normalized to the highest measured dose rate arising from
neutrons. At this point, the absolute data values (in terms of
ambient dose equivalent) are under scrutiny. The detectors used are
calibrated for a continuous radiation emission and not for radiation
produced by a machine running at 10~Hz. The top plot, in
Fig.\ref{figRadMeasur}, shows the dose rate produced when the
machine is accelerating the electron beam produced by the laser to
230~MeV and the dark current. The bottom plot shows solely the
radiation produced by the dark current (laser is OFF). The
contribution to the dose rate by the losses of electrons from the
SwissFEL electron beam is negligible compared to the dark current
produced by the gun. In addition, and not shown here, when the gun
RF and the laser are turned off, the dose rate measured is close to
background in most sections of the machine and of a few percent near
the BC.

In order to reduce the radiation dose produced by lost electrons
from the dark current, we have inserted an aperture at the location
of the first screen. It is located downstream of the first WCM and
before the first BPM. The second WCM, downstream of the BC, does not
record any dark current when this aperture is introduced in the beam
line. Concurrent radiation dose measurements are currently under
analysis.

Overall, on-line radiation and dark current measurements are in
qualitative agreement with Elegant simulation results performed
using SwissFEL's injector layout \cite{SwissFEL:CDR}.

\subsection{Dark current of a C-band test structure}

\begin{figure}[htb]
   \centering
   \includegraphics*[width=\columnwidth, clip=]{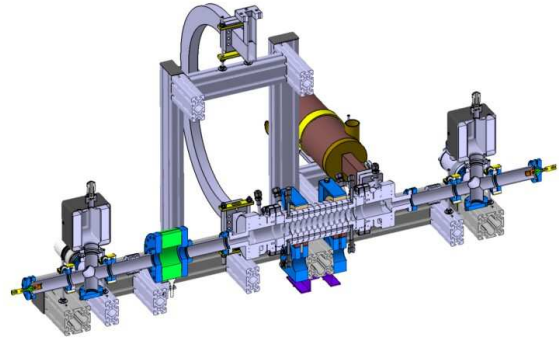}
   \caption{C-Band TW structure test stand, showing a faraday cup at each end of the beam line and
   an ICT.}
   \label{figCbandStand}
\end{figure}

The main acceleration of SwissFEL will be provided by C-band
structures with an RF klystron pulse of 350~ns flat top and an
on-axis maximum gradient of 28~MV/m \cite{SwissFEL:CDR}. It is of
importance to test not only the quality of the mechanical production
of such structures, but also their RF properties
\cite{Zennaro:IPAC2012}. Two small prototypes have been RF tested,
with a repetition rate of 10~Hz for the first structure and 100~Hz
for the second one. Their dark currents were measured using the two
Faraday cups (FCs) installed at each end of the beamline
Fig.\ref{figCbandStand}. For the future structures an Integrated
Current Transformer (ICT) \cite{Bergoz} will be installed in the
beamline.

\begin{figure}[htb]
   \centering
   \includegraphics*[width=\columnwidth, clip=]{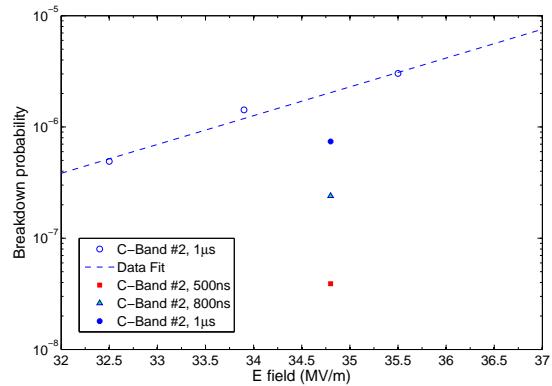}
   \caption{Breakdown probability of a C-band structure running at 100~Hz repetition rate
   for various RF pulse length and accelerating gradient.}
   \label{figCbandBDR}
\end{figure}

We did not record any dark current on either the upstream or
downstream FCs while running the C-band cavity for many hours at an
accelerating gradient of 38~MV/m, hence 10~MV/m above the design
value, and with a 350~ns RF flat top pulse. This amounts to
$\sim$51~MW of input power. The second C-band structure is currently
RF tested with a flat top pulse already exceeding 500~ns. In both
cases, dark current was detected solely on the Faraday cups during
breakdowns. The breakdown rate is shown, Fig.\ref{figCbandBDR}, for
the second C-band structure for various accelerating field and pulse
length. The data were obtained after various conditioning hours at
different pulse length and RF power hence the discrepancy observed
at 1~$\mu$s. The lowest point at 1~$\mu$s has been obtained before
the dismantling of the test Structure \#2. This shows the effect of
a good conditioning.

\begin{figure}[htb]
   \centering
   \includegraphics*[width=\columnwidth, clip=]{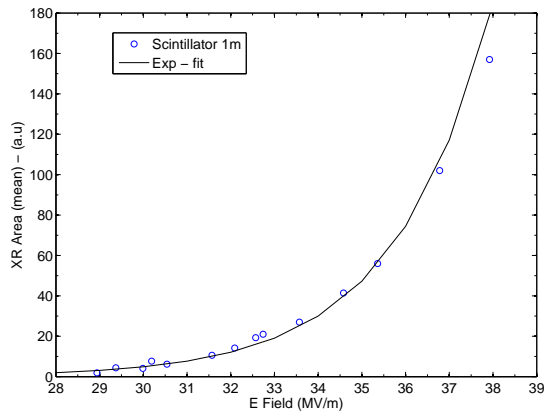}
   \caption{X-ray signal in function of the Electric Field inside the C-Band TW structure.
    The curve shows a FN type electron emission.}
   \label{figCbandXray}
\end{figure}

The use of an X-ray scintillator \cite{lepimpec:IPAC2010}, placed
1~m away from the structure, proves that dark current exists inside
the structure. Using the NIST \cite{NIST} database, we determined
that photons $>$ 300~keV can make it through 40~mm of Cu. At 38~MV/m
on axis field, an electron in 1 cell (2~cm) can acquire up to
750~keV. The X-ray data are well fitted by the use of an exponential
function, as expected for dark current coming from field emission,
Fig.\ref{figCbandXray}.

Using equation~\ref{EquFNbeta} and assuming that the integrated
x-ray signal is proportional to the dark current, we estimated that
the FN $\beta$ is 68 (350~ns long pulse). This value is consistent
for the two structures tested. The evolution of the FN $\beta$ was
also recorded after a major breakdown. $\beta$ peaked to 150 and
diminished to 65 after hours of conditioning.

\section{CONCLUSIONS}

In order to obtain the necessary official authorization to operate
SwissFEL, the production of radiation along the machine must be
quantified. We have produced a first set of simulation results using
ASTRA, Elegant and OPAL as well as carrying out measurements at two
SwissFEL facilities. The outcomes are :

\begin{itemize}
 \item[*] Like in many other facilities around the world, the dark current is
  essentially produced by the SW photogun.
 \item[*] From simulations: most of the dark current is lost at the first S-band structure.
 \item[*] The remainder of it is mostly lost before the first bunch
 compressor.
 \item[*] Qualitative and comparative online radiation measurements show a low dose rate
 at the first S-band structure (unexpected), a high dose rate at the fourth S-band structure (unexpected)
 and a high dose rate at the entrance of the BC (expected)
 \item[*] Only a small fraction of the dark current, mainly its core produced at the
 photocathode, goes to the end of SwissFEL.
 \item[*] The S-band structures do not produce any recordable dark current nor
 X-rays when running at nominal power.
 \item[*] At nominal power and pulse length, the C-band test
 structures produce no measurable dark current, although electrons are present in the structures (X-ray signal).
\end{itemize}

Finally, we are continuing the dark current investigation on
subsequent C-band structures by installing an ICT closer to the
exit/entrance of the structure. We will run OPAL using the C-band
test structure geometry and use the X-ray data to benchmark the
simulations. Following the radiation measurements and simulations
results, the SwissFEL layout infrastructure (air vents and
shielding) is undergoing some modifications.

\section{ACKNOWLEDEGMENTS}

We would like to thank A. Jurgens for his help at the C-band test
facility. The lead author is indebted to his beam dynamic colleagues
for their various help. We also acknowledge the help of C. Wang
(CIAE, China) for modifying OPAL according to our needs. We also
thank the radiation protection group for their support of this work.

%
%


\end{document}